# Re-appraisal and extension of the Gratton-Vargas two-dimensional analytical snowplow model of plasma focus - Part II: Looking at the singularity

S K H Auluck

HiQ TechKnowWorks Private Limited, Nerul, Navi Mumbai, 400706 India

email: skhauluck@gmail.com

Abstract

The Gratton-Vargas snowplow model, recently revisited and expanded (S K H Auluck, Physics of Plasmas, 20, 112501 (2013)), has given rise to significant new insights into some aspects of the Dense Plasma Focus (DPF), in spite of being a purely kinematic description having no reference to plasma phenomena. It is able to provide a good fit to the experimental current waveforms in at least 4 large facilities. It has been used for construction of a local curvilinear frame of reference, in which conservation laws for mass, momentum and energy can be reduced to effectively-one-dimensional hyperbolic conservation law equations. Its utility in global parameter optimization of device parameters has been demonstrated. These features suggest that the Gratton-Vargas model deserves a closer look at its supposed limitations near the singular phase of the DPF. This paper presents a discussion of its development near the device axis, based on the original work of Gratton and Vargas, with some differences. It is shown that the Gratton-Vargas partial differential equation has solutions for times after the current singularity, which exhibit an expanding bounded volume, (which can serve as model of an expanding plasma column) and decreasing dynamic inductance of the discharge, in spite of having no built-in hydrodynamics. This enables the model to qualitatively reproduce the characteristic shape of the current derivative in DPF experiments without reference to any plasma phenomena such as instabilities, anomalous resistance or reflection of hydrodynamic shock wave from the axis. The axial propagation of the solution exhibits a power-law dependence on the dimensionless time starting from the time of singularity, which is similar to the power-law relations predicted by theory of point explosions in ideal gases and which has also been observed experimentally.

**Introduction:**

The Gratton-Vargas (GV) model[1] of the Dense Plasma Focus (DPF) is a purely kinematical model. It is based on the so called "snowplow hypothesis": that the shape and speed of the DPF plasma current sheath (PCS) is governed by a balance between the magnetic pressure driving the PCS into the neutral gas ahead of it and the "wind pressure" resisting its motion. This assumption, along with the assumption of azimuthal symmetry, allows formulation of a partial differential equation for an imaginary mean surface (referred henceforth as the GV Surface or GVS) in the cylindrical (r,z) space connecting inner and outer electrodes, representing the shape and position of the PCS. This imaginary surface has no attributes of a plasma (such as density, temperature, fluid velocity and their spatial and temporal distributions) associated with it; rather, it has the same utility as the concept of center-of-mass in mechanics.

The pioneering contribution of F. Gratton and J. M. Vargas [1] lies in construction of this partial differential equation, discovery of its analytical solutions and some of their properties. Their crucial innovation consists of definition of a dimensionless independent variable $\tau$ which plays the role of time and which is defined in terms of the charge that has flown in the circuit in time t, normalized to a "mechanical equivalent of charge" defined in terms of device parameters. This allows determination of the shape and location of the PCS (or rather its representation in terms of the idealized GV surface) at various values of $\tau$, from which the dynamic inductance attributable to the moving PCS can be determined. The circuit equation can be easily solved in terms of a tabulated functional dependence of dynamic inductance $\mathcal{L}(\tau)$, which can be empirically fitted to simple algebraic expressions. The solution for current as function of $\tau$ is used to recover real time t and thus the actual current profile can be calculated.

The GV model was developed in an age when computing power was expensive and not easily accessible. So Gratton and Vargas tried to rely on analytical techniques as much as possible and avoided computationally intensive investigations. One of their assumptions was the neglect of circuit resistance, because of which, their model was inherently incapable of providing quantitative fits to experimental current waveforms. This deficiency was removed recently [2], by using a successive approximation method to take into account the circuit resistance. As a result, it has become possible to demonstrate good fit to experimental current waveforms for 4 large experimental facilities[3] using a judicious choice of static inductance, circuit resistance and gas pressure as fitting parameters. The analytically-defined shape of the imaginary GV surface representing PCS has been used [4] to construct a local curvilinear coordinate system attached with the GVS, with unit vectors tangent to GVS, along the azimuth and along the normal to GVS. In this coordinate system, conservation laws for mass, momentum and energy can be reduced to effectively-one-dimensional hyperbolic conservation law form with geometric and time-dependent source terms. This construction has been used to demonstrate that axial magnetic field and toroidally moving fast ions, which have been inferred experimentally over three decades of research, are natural consequences of conservation laws in the curved axisymmetric geometry of the DPF current sheath [4]. An empirical numerical formula for the dynamic inductance of PCS as a function of dimensionless time $\tau$, fitted to thousands of numerical calculations, has been used to illustrate the possibility of global optimization of device parameters with respect to arbitrarily selectable performance criteria [5].

With so much going in favor of the GV model, it is useful to look at its limitations. Its purely kinematic nature precludes any predictions concerning formation of plasma column (its radius, shape, density or temperature). This is in stark contrast with the popular and powerful Lee model [6], which has been used to look at scaling laws for various emissions from the plasma. The fundamental assumption of snowplow model is expected to fail[2] when the PCS reaches close to the axis, where the gas dynamic shock would get reflected from the axis and stop the inward movement of the PCS, causing it to stagnate, resulting in a plasma column of finite radius. Because of this, all discussions [2-5] of the revised resistive GV model have so far been restricted artificially to the stage where the intersection of the GV surface with the flat-top anode, reaches an empirically derived[7] radius of 0.12 times the anode radius - considering the region beyond that stage as *terra incognita*.

This paper continues the discussion [2] on the Gratton-Vargas snowplow model of the Dense Plasma Focus to the earlier-restricted zone near the device axis and the current singularity. It must be emphasized that this is an investigation of a mathematical model: *not of a plasma*. The objective is to compare what this admittedly-oversimplified [2] model predicts with what is actually observed. The motivation for this exercise lies in the expectation that discovery of any close correspondence between predictions and observations should reveal a weak dependence of some phenomena on the details of plasma processes (which would be an important insight), whereas discrepancies should reveal clues to construction of better first-principles theories, such as those based on hyperbolic conservation laws [4] in the local curvilinear coordinate system.

The discussion presented in this paper implicitly relies on the fact that the open-ended plasma configuration in the so called "pinch phase" of DPF cannot create a pressure equilibrium as assumed in the infinite-length Bennett z-pinch concept [8] - the *pressure must relax by axial outflow*. Since the collapse occurs at the anode surface first and then the point of contact "zippers up" along the axis, the condition assumed in the snowplow hypothesis - neutral gas on one side and fully ionized current-carrying plasma on the other side - may be expected to continue to hold during this process. Thus, it should be possible to describe an axially moving snowplow shock, simulating the "zipper effect", using the GV formalism. The original paper of Gratton and Vargas [1] does provide a recipe for constructing such shock; however, this was never followed up probably because of the computational difficulties involved. In fact, the only necessary condition for the validity of the snowplow hypothesis is the requirement that the magnetic field and the fluid move together as a single shock front; considerations of conservation laws and ionization stability presented elsewhere [4,9] reveal that this is possible in a restricted range of plasma parameters, which remains to be fully explored. Until then, the existence of such range of parameters can be taken as a working hypothesis to assert that the GV model is not necessarily and inherently restricted to the region beyond the immediate vicinity of the axis and times earlier than the current derivative singularity.

The next section provides a brief overview of the earlier theoretical development for the sake of continuity, with emphasis on the geometric and technical aspects of the solution. Section III describes the GV construction of the axial shock and its numerical illustration. Section 4 describes calculation of the inductance beyond the singularity and shows that the GV model is capable of qualitatively reproducing the characteristic current and current derivative signatures of DPF experiments. It also demonstrates that the axial movement of the solution obeys a power-law dependence on the dimensionless time. Section 5 presents summary and conclusions.

# I. Overview of the GV model:

This section attempts a brief overview of the GV model for the sake of providing a starting point for the present discussion while avoiding unnecessary repetition; however, a re-read of the relevant papers [2,4,5] would be greatly useful in appreciating the foregoing arguments and understanding the nomenclature used. While these earlier papers focused on the conceptual foundations of the GV model, the present overview focuses on the technical aspects of constructing acceptable solutions of the GV partial differential equation given below:

$$\partial_\tau \psi + \sqrt{(\partial_{\tilde{r}}\psi)^2 + (\partial_{\tilde{z}}\psi)^2}\, \frac{1}{2\tilde{r}} = 0 \qquad 1$$

This equation describes the evolution of an imaginary mean surface $\psi(r,z,t) \equiv z - \overline{f}(r,t) = 0$ representing the position and shape of the plasma current sheath (PCS) in the two dimensional space defined by dimensionless coordinates $\tilde{r} \equiv r/a$ and $\tilde{z} \equiv z/a$ as function of an independent variable $\tau$ defined as

$$\tau(t) = \frac{1}{Q_m} \int_0^t I(t')\, dt', \quad Q_m \equiv \frac{\pi a^2 \sqrt{2\mu_0 \rho_0}}{\mu_0} \qquad 2$$

where a is the anode radius, $\rho_0$ is mass density of the fill gas, $I(t)$ is the temporal current profile and $Q_m$ is a "mechanical equivalent of charge".

A family of general solutions of equation 1, corresponding to initial condition specified by an "initial PCS profile" $\tilde{z}_i = \overline{f}_i(\tilde{r}_i, \tau_i)$ at initial "instant" $\tau_i$, can be constructed by the method of characteristics as discussed elsewhere [1,2] in detail. The set of characteristics everywhere normal to the integral surface representing the solution of 1 is given by equation

$$\frac{\tilde{z}}{N} + s\,\text{ArcCosh}\left(\frac{\tilde{r}}{|N|}\right) = C_1 = \text{Constant} \qquad 3$$

Here, N is an invariant defined by the method of characteristics applied[2] to equation 1, which labels the characteristics. The symbol s stands for the sign of $\left(d\bar{f}_i(\tilde{r}_i,\tau_i)/d\tilde{r}_i\right)$ in its domain of definition. The location of the GV surface - the integral surface of partial differential equation 1- along the characteristic curve 3 at time $\tau$ is given[2] by the equation

$$\frac{\tilde{r}}{N}\sqrt{\frac{\tilde{r}^2}{N^2}-1} + \text{ArcCosh}\left(\frac{\tilde{r}}{|N|}\right) + \frac{s\tau}{N^2} = C_2 = \text{Constant} \qquad 4$$

The initial profile $\tilde{z}_i = \bar{f}_i(\tilde{r}_i,\tau_i)$ provides the values of the constants $C_1$ and $C_2$ in terms of $N_i(\tilde{r}_i,\tilde{z}_i;\tau_i) = \tilde{r}_i\cos\phi_i \neq 0$ where $\phi_i$ is the angle made by the normal to the initial profile at $(\tilde{r}_i,\tilde{z}_{i,})$ with the z-axis:

$$C_1(\tilde{r}_i,\tilde{z}_{i,};N_i) = \frac{\tilde{z}_i}{N_i} + s\text{ArcCosh}\left(\frac{\tilde{r}_i}{|N_i|}\right) \qquad 5$$

$$C_2(\tilde{r}_i,\tau_i,N_i,s) = \frac{\tilde{r}_i}{N_i}\sqrt{\frac{\tilde{r}_i^2}{N_i^2}-1} + \text{ArcCosh}\left(\frac{\tilde{r}_i}{|N_i|}\right) + \frac{s\tau_i}{N_i^2} \qquad 6$$

The substitution $\alpha/2 \equiv sC_1(\tilde{r}_i,\tilde{z}_{i,};N_i) - s\tilde{z}/N_i$, $\tilde{r} = |N|\text{Cosh}(\alpha/2); \tilde{z} \equiv N(C_1 - s\alpha/2)$ is seen to satisfy 3 identically. The same substitution in 4 leads to

$$F(\alpha) \equiv \text{Sinh}(\alpha) + \alpha = 2\left(C_2(\tilde{r}_i,\tau_i,N_i) - \frac{s\tau}{N_i^2}\right) \qquad 7$$

In order to actually compute the family of characteristic curves labeled by $N_i$ and the orthogonal family of GV surfaces (which are curves in the 2D r-z space) labeled by $\alpha$, which is determined as a function of $\tau$ using 7, one needs to define the geometry of the electrodes and the boundary conditions. The GV model is, in principle, general enough to describe the snowplow discharge between an arbitrary pair of electrodes. For the case of Mather type DPF, the geometry consists of a pair of coaxial electrodes and a plasma-formation region represented by the insulator, shown in Fig. 1.

The boundary condition consistent with the idea of representing the shape and position of PCS with an imaginary surface of zero thickness is that the GV surface must meet the anode at a right angle[2]. This implies that the characteristics must emerge in all directions from a vertex of the anode profile in (r,z) space and that they must be tangent to the straight vertical edges of the anode profile. This is illustrated in Fig. 1(a).

Along the insulator length, the characteristics are straight lines normal to insulator surface: this case of N=0 is not covered by the solution 3-6 and needs to be dealt with separately[2]. At vertex A at the top of the insulator, a continuum of characteristics is emitted with $N_i$ changing from 0 to $\tilde{r}_i$, a few of which are shown. Starting with the junction of insulator and anode, a series of characteristics with $N_i=1$ and s=-1 are emitted from each point on the anode until the top vertex B. Characteristics starting from B with $N_i$ varying from 1 (thick line from the vertex, green in the online version) to zero and continuing to negative values are shown. Note that these characteristics are bounded by an envelope (shown by thick curve colored blue in the online version) which is tangent to the axis and to each characteristic. The slope of the characteristics changes sign at the thick dashed closed line. Characteristics are also shown in the region below the top of the anode, where $N_i$ has negative values: if there is a cavity in the anode, the solution can be continued into the cavity using the downward-going characteristics.

Also shown in the region between the envelope and the axis are a set of thick, dashed (blue in the online version) curves. These are the upper branch of the envelope repeated along the axis. They are discussed in the next section.

Technically, computation of each characteristic curve involves the parametric function $\tilde{r} = \tilde{r}(\alpha, N_i); \tilde{z} = \tilde{z}(\alpha, N_i, s, \tilde{r}_i, \tilde{z}_i)$. In each region, $(\tilde{r}_i, \tilde{z}_i)$ is either a specified point (a vertex) with $N_i$ taking a continuous range of values (0 to $\tilde{r}_i$ at the vertex at the top of insulator or 1 to 0 at the end of anode) or a set of points (for example, along the cylindrical anode surface) with a fixed value for $N_i$ (1 in the rundown region). For these sets of values, solution of the simultaneous equations $\tilde{r} = \tilde{r}_i = |N_i|\text{Cosh}(\alpha/2); \tilde{z} = \tilde{z}_{i,} = N_i\left(C_1(\tilde{r}_i, \tilde{z}_{i,}; N_i) - s\alpha/2\right)$ provides a set of values of $\alpha$, using which the parametric function $\tilde{r} = \tilde{r}(\alpha, N_i); \tilde{z} = \tilde{z}(\alpha, N_i, s, \tilde{r}_i, \tilde{z}_{i,})$ can be plotted. Both values s=+1 and s=-1 are used in calculating the characteristics from the vertex B at the end of anode. In this case, $N_i$ is varied from +1 to -1; the negative values generate characteristics below the anode.

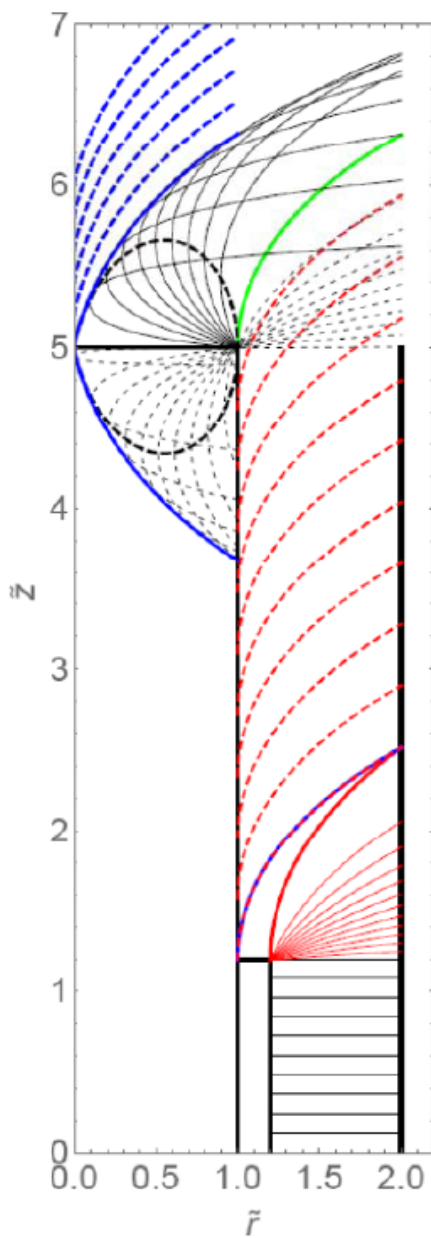
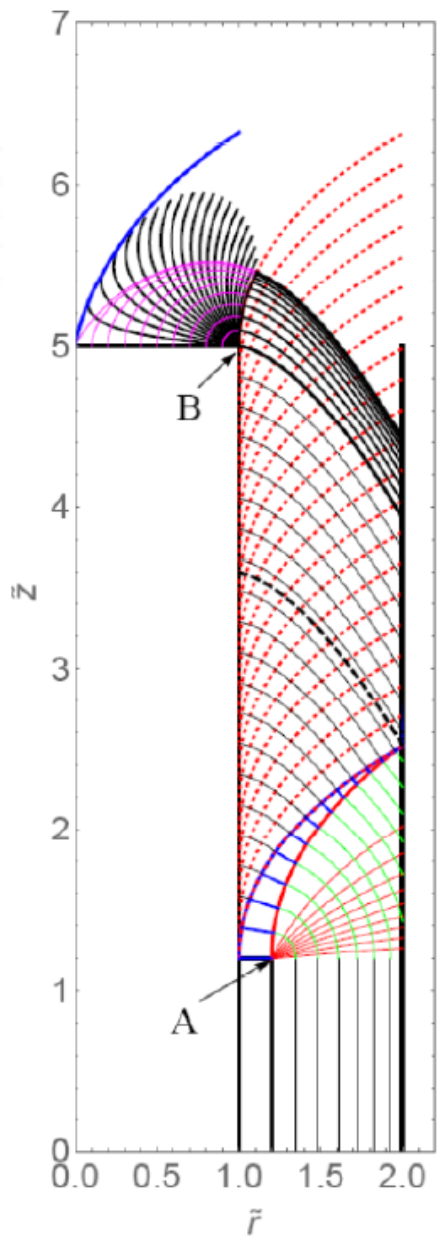

Fig 1(a)     Fig 1(b)

Fig. 1: Illustration of the characteristics and the solution of GV partial differential equation for Mather type DPF. For the sake of illustration, the following geometrical parameters of DPF normalized to the anode radius are chosen: anode length $\tilde{z}_A = 5$, cathode radius $\tilde{r}_C = 2$, insulator radius $\tilde{r}_I = 1.2$, insulator length $\tilde{z}_I = 1.2$. For details, see text.

Fig 1(b) illustrates the computation of the GV surface. Again the parametric function $\tilde{r} = \tilde{r}(\alpha, N_i); \tilde{z} = \tilde{z}(\alpha, N_i, s, \tilde{r}_i, \tilde{z}_i)$ is plotted, but the value of $\alpha$ is obtained from 7 by evaluating the inverse function $F^{-1}(\text{Sinh}(x) + x)$. GV provide [1] the following inversion formula:

$$\alpha = \frac{F}{2}\left(1 - \frac{F^2}{48}\right) \quad \text{for} \quad |F| < 1$$

$$\alpha = \text{Log}\left[2F + \frac{1}{2F} - 2\left(1 - \frac{1}{F}\right)\text{Log}(2F)\right] \quad \text{for} \quad |F| \gg 1$$

The value of F is found from the right hand side of 7. This makes the value of $\alpha$ dependent on $\tau$ and on the initial value of radius $\tilde{r}_i$ which is a fixed value at the vertices. In the rundown region, equations 3 and 4 reduce to a travelling-wave form[2] using s=-1 and N=1, eliminating the need for determining $\alpha$ by a computationally intensive procedure. At the vertex at the end of anode, the initial radius is reset to the normalized anode radius 1, the initial value of $\tau$ is reset to its value $\tau_r = 2(\tilde{z}_A - \tilde{z}_I)$ at the end of rundown, the value of s changes from -1 during rundown to +1. When the parametric function $\tilde{r} = \tilde{r}(\alpha, N_i); \tilde{z} = \tilde{z}(\alpha, N_i, s, \tilde{r}_i, \tilde{z}_i)$ is evaluated using values of $\alpha$ determined from $F^{-1}$ (or using the travelling wave solution), the result is a family of curves which resemble the PCS evolution. This is illustrated in Fig 1(b), which also shows corresponding characteristic curves for comparison .

Note that the GV surface touches the flat top of the anode at right angles, satisfying the boundary condition. The maximum value of $\tau$ that yields a solution satisfying this boundary condition is $\tau_r + 1$. Beyond that, the solution touches the envelope orthogonally, a phenomenon discussed in the next section. It also becomes clear that *the space between the axis and the envelope cannot be approached by any solution constructed using characteristics emitted from the vertex at the top of anode*.

Determination of the shape of the GV surface, which is taken to represent PCS position and shape, enables calculation of the total flux due the azimuthal magnetic field contained between the GV surface and the electrodes, which leads to numerical evaluation of the functional dependence of dynamic inductance $\mathfrak{L}(\tau)$ of the discharge, from which the circuit equation can be solved by the method described elsewhere[2]. The dynamic inductance tends to increase very sharply as the calculation approaches the device axis closely, leading to division by zero error at $\tau = \tau_r + 1$. The limitation of the GV model (as presently elaborated) in dealing with the singular phase arises solely from the fact the solution to GV equation found from the method of characteristics using the procedure described above results in values of $\mathfrak{L}(\tau)$ up to but excluding $\tau_r + 1$. There is no mechanism, similar to the slug model hypothesis incorporated in the Lee model[6], which can make the GV surface representing the PCS move radially outward, leading to decreasing dynamic inductance, which should be responsible for the slight increase of current after the singularity. The present version of GV model therefore leads up to the beginning of current dip, but not to times after the dip, so that the depth of the current dip and associated signatures in voltage and dI/dt are not covered by the model as it stands. The next section discusses the prescription suggested by Gratton and Vargas[1] and how it works in practice.

## II. **GV model near the axis and at the singularity:**

The key to the development of the GV model in the vicinity of the axis (and at the current singularity) is the observation by Gratton and Varga[1] that the envelope of the characteristics from the vertex at the end of anode is composed of infinitesimal pieces of an infinite number of characteristics and *thus must itself be a characteristic* which should be normal to a proper solution of the GV equation. The envelope, a curve that is tangent to each member of a family of curves, is obtained by the following procedure.

For the family of characteristic curves $\tilde{r} = \tilde{r}(\alpha, N_i); \tilde{z} = \tilde{z}(\alpha, N_i, s, \tilde{r}_i, \tilde{z}_i)$ with s=-1 and $\tilde{r}_i = 1$, $\tilde{z}_i = \tilde{z}_A$, the following equation is solved numerically for α>0 with $N_i$ varying from a very small value approximating zero to unity

$$\frac{\partial \tilde{r}(\alpha, N_i)}{\partial \alpha}\frac{\partial \tilde{z}(\alpha, N_i, s, \tilde{r}_i, \tilde{z}_i)}{\partial N_i} - \frac{\partial \tilde{r}(\alpha, N_i)}{\partial N_i}\frac{\partial \tilde{z}(\alpha, N_i, s, \tilde{r}_i, \tilde{z}_i)}{\partial \alpha} = 0 \qquad 9$$

The resulting tabulated function $\alpha(N_i)$ is used along with the parametric equation $\tilde{r} = \tilde{r}(\alpha(N_i), N_i); \tilde{z} = \tilde{z}(\alpha(N_i), N_i, s, \tilde{r}_i, \tilde{z}_i)$ to generate the branch of the envelope above the anode. This has a reasonably good fit to the following expression for $0 < \tilde{r} < 1$ (Fig 3)::

$$\tilde{z} = \tilde{z}_i + a\tilde{r}^b; \quad a = 1.36307, b = 0.603254 \qquad 10$$

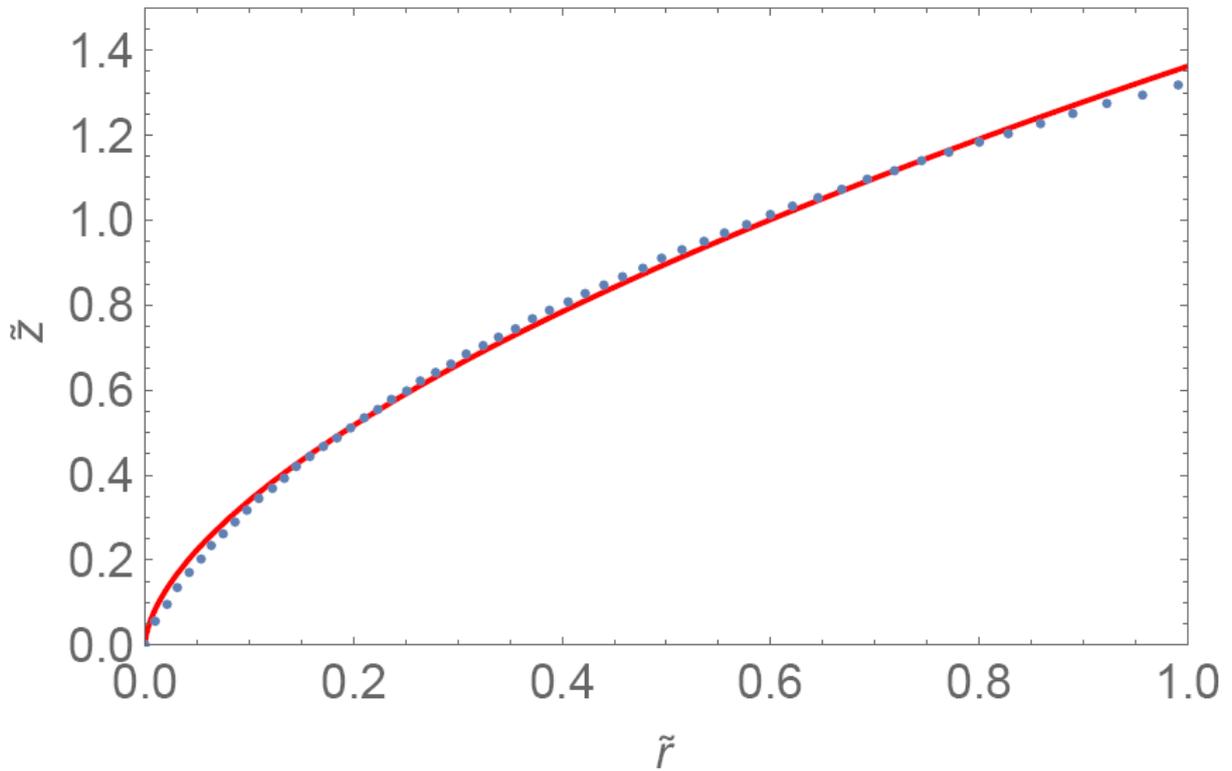

Fig 2. The dots represent the numerically calculated envelope, the solid line (red in online version) is the fitted curve 10.

The only free parameter of this curve is $\tilde{z}_i$. Thus a family of such curves, displaced along the axis can be created by treating $\tilde{z}_i$ as a parameter, as illustrated in Fig. 1(a). The orthogonal trajectories of this family, which must be a solution of the GV equation according to Gratton and Vargas[1], are easily found as

$$\tilde{z} = c - \frac{1}{ab(2-b)} \tilde{r}^{2-b}$$

The constant c is chosen in such manner that the orthogonal curve - expected (according to Gratton and Vargas [1]) to be a solution of the GV equation 1 - meets the envelope at specified points $(\tilde{r}_{env}, \tilde{z}_{env})$:

$$c = \tilde{z}_{env} + \frac{1}{ab(2-b)} \tilde{r}_{env}^{2-b} \qquad 11$$

The value of $\tau_{env}$ at which the portion of GV surface, determined using the procedure described in the previous section, meets the envelope at the specified points $(\tilde{r}_{env}, \tilde{z}_{env})$ is found by equating corresponding $\alpha(N_i)$ found from the solutions of 9 to $\alpha$ given by 7, using suitable algorithm for inversion of $F(\alpha)$, with $\tilde{r}_i = 1, \tau_i = \tau_r$, s=+1, $\tau = \tau_{env}$. Fig 3 illustrates the result.

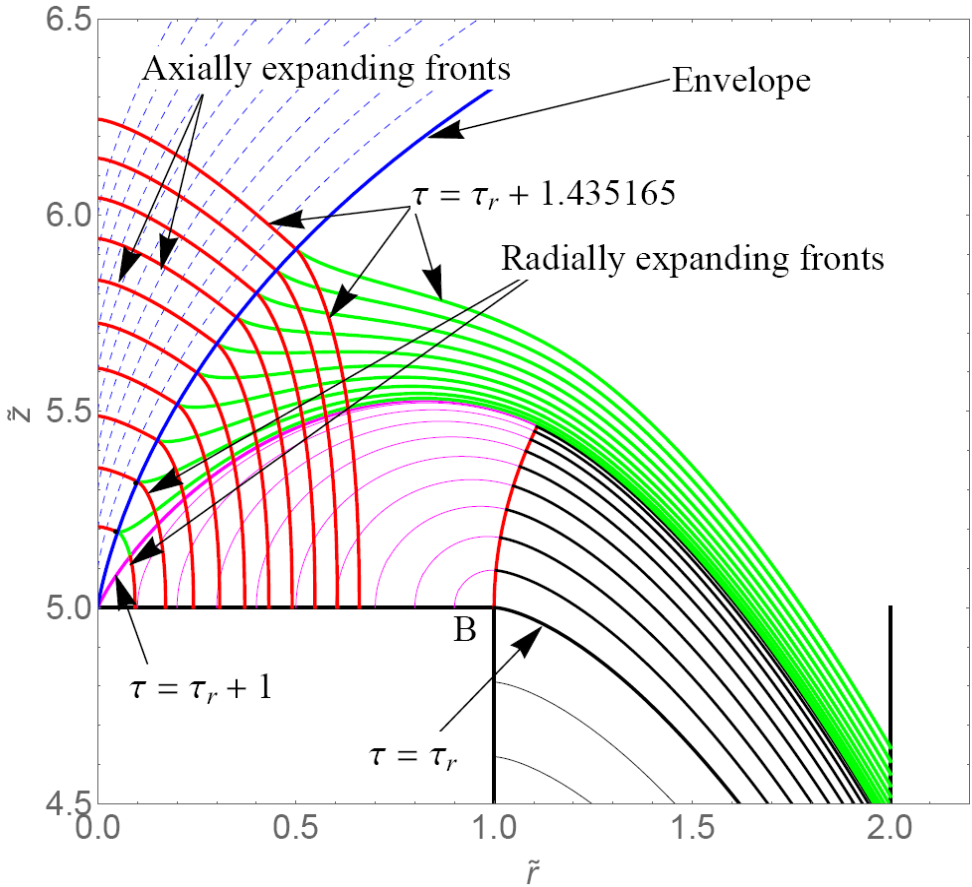

Fig. 3: Gratton-Vargas construction of the solution between the axis and the envelope. See text for details.

The dashed lines (blue in online version) between the axis and the envelope are copies of the envelope translated along the axis, providing a family of characteristics, whose orthogonal trajectories, (shown in thick red lines in the online version) are solutions to the GV equation according to the construction proposed by Gratton and Vargas [1]. Between the axis and the envelope, they represent axially moving fronts. At their intersection with the envelope, each of them forms two branches: one goes from the point of intersection with the envelope to meet the anode orthogonally, satisfying the boundary condition [2]; the other goes on to meet the travelling wave solution which connects it to the cathode. It would be reasonable to suppose that the radially and axially expanding fronts together form a closed expanding volume, on whose boundary, current could flow from the anode to the cathode, first along the radially expanding front to its intersection with the envelope and then onwards to the cathode along the other branch.

This provides an algorithmic basis for calculating the flux of the azimuthal magnetic field and hence the expected dynamic inductance of the discharge in the phase after the singularity; this is explored in the next section.

### III. **Calculation of dynamic inductance after the GV surface reaches axis:**

One attractive feature of the GV model is that the solution of the GV equation after the rundown phase $(\tau = \tau_r)$ is completely independent of device parameters and device type (Mather, Filippov or hybrid). This allows a once-for-all tabulation of data related with the solution; in the present context this particularly applies to the envelope of characteristics emitted at the vertex B at the end of anode and associated numbers. Table I summarizes such data.

The data refers to the ten solutions beyond $\tau_r + 1$ shown in Fig. 3. The "volume" refers to the dimensionless volume enclosed between the radial and axial expanding fronts and the anode. For evenly spaced values of $\tilde{r}_{env}$, the table provides values of $N_{env}$ and $\alpha_{env}$ calculated from 9 and resulting values of $\tilde{z}_{env} - \tilde{z}_A$ and $\tau - (\tau_r + 1)$. The previous-to-last column provides the radius of the radially expanding front at the anode surface. The last column provides the axial position of the front at the axis. Fig. 4 shows comparison of the axial position of the shock with a power-law fit.

**Table 1: Some numerical data regarding envelope**

| $\tilde{r}_{env}$ | $\tilde{z}_{env} - \tilde{z}_A$ | $N_{env}$ | $-10^2 \alpha_{env}$ | $\tau - (\tau_r + 1)$ | volume | $\tilde{r}_{front}$ | $\tilde{z}_{front}$ |
|---|---|---|---|---|---|---|---|
| 0.05 | 0.1930 | 0.0530 | 0.8601 | 0.0088 | 0.001094 | 0.0938 | 0.206 |
| 0.10 | 0.3212 | 0.1098 | 0.8096 | 0.02923 | 0.007379 | 0.1710 | 0.356 |
| 0.15 | 0.4265 | 0.1707 | 0.7591 | 0.05824 | 0.02225 | 0.2413 | 0.488 |
| 0.20 | 0.5177 | 0.2367 | 0.7086 | 0.09453 | 0.04838 | 0.3075 | 0.610 |
| 0.25 | 0.5986 | 0.3088 | 0.6581 | 0.13732 | 0.08798 | 0.3706 | 0.724 |
| 0.30 | 0.6719 | 0.3883 | 0.6076 | 0.18608 | 0.1430 | 0.4314 | 0.834 |
| 0.35 | 0.7391 | 0.4771 | 0.5570 | 0.24047 | 0.2152 | 0.4904 | 0.940 |
| 0.40 | 0.8011 | 0.5776 | 0.5065 | 0.30023 | 0.3060 | 0.5479 | 1.043 |
| 0.45 | 0.8589 | 0.6932 | 0.4559 | 0.36517 | 0.4170 | 0.6043 | 1.144 |
| 0.50 | 0.9131 | 0.8287 | 0.4053 | 0.43516 | 0.5493 | 0.6597 | 1.244 |

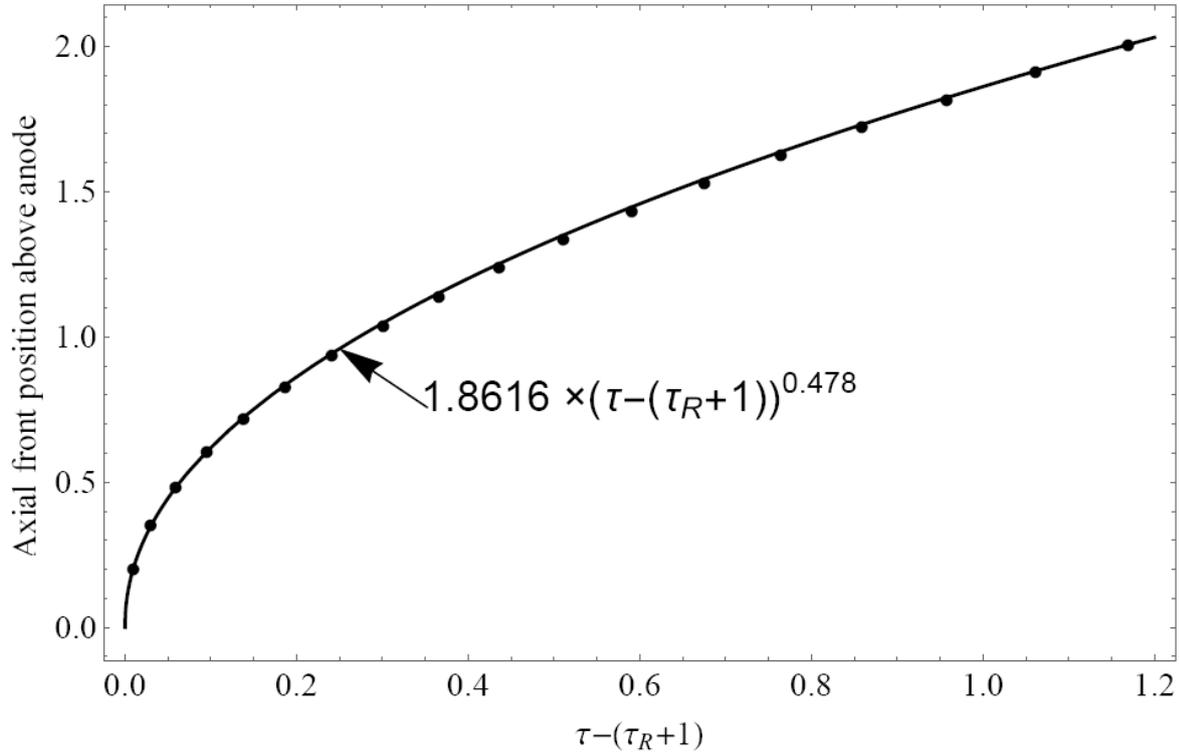

Fig. 4: Comparison of the axial front position above the anode with a power-law fit.

The dynamic inductance of a solution with $\tau$ values on either side of $\tau_r + 1$ is calculated by evaluating the flux integral

$$\mathcal{L}(\tau) = \iint d\tilde{z}d\tilde{r}\frac{1}{\tilde{r}} \qquad 12$$

over the domain bounded by the solution and the electrode system as illustrated in Fig. 5. The calculated inductance variation is shown in Fig. 6. This calculation uses values of $\tau$ reaching within $10^{-5}$ of $\tau_r + 1$, while excluding this singular value. The resulting current normalized to $I_0$, and current derivative normalized to $V_0/L_0$ are shown in Fig. 7 and 8 as functions of time normalized to the quarter-cycle time period, where the current derivative is calculated from

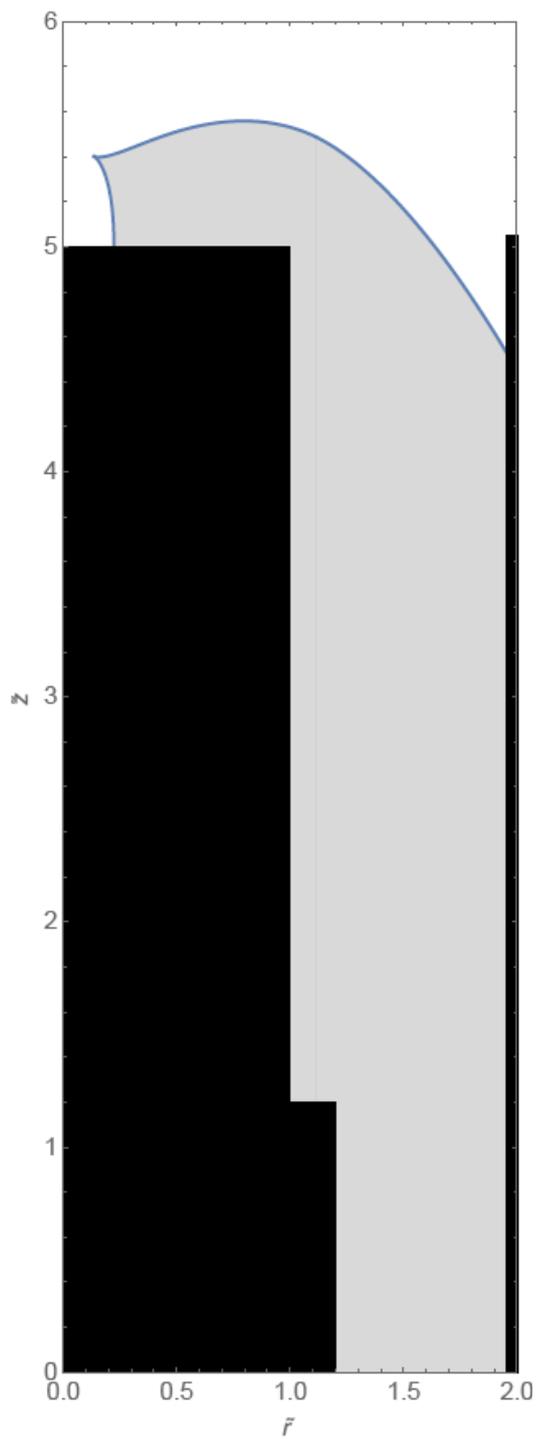

$$\frac{L_0}{V_0}\frac{dI}{dt} = \frac{1}{2\varepsilon}\frac{d\tilde{I}^2}{d\tau}$$

For this demonstration, the value of $\varepsilon$ was chosen in such way that the energy remaining in the capacitor[2] at $\tau_r + 1$ is zero: $\varepsilon \equiv Q_m/C_0 V_0 = (\tau_r + 1)^{-1}$.

Fig. 5: The gray shaded area illustrates the domain of integration

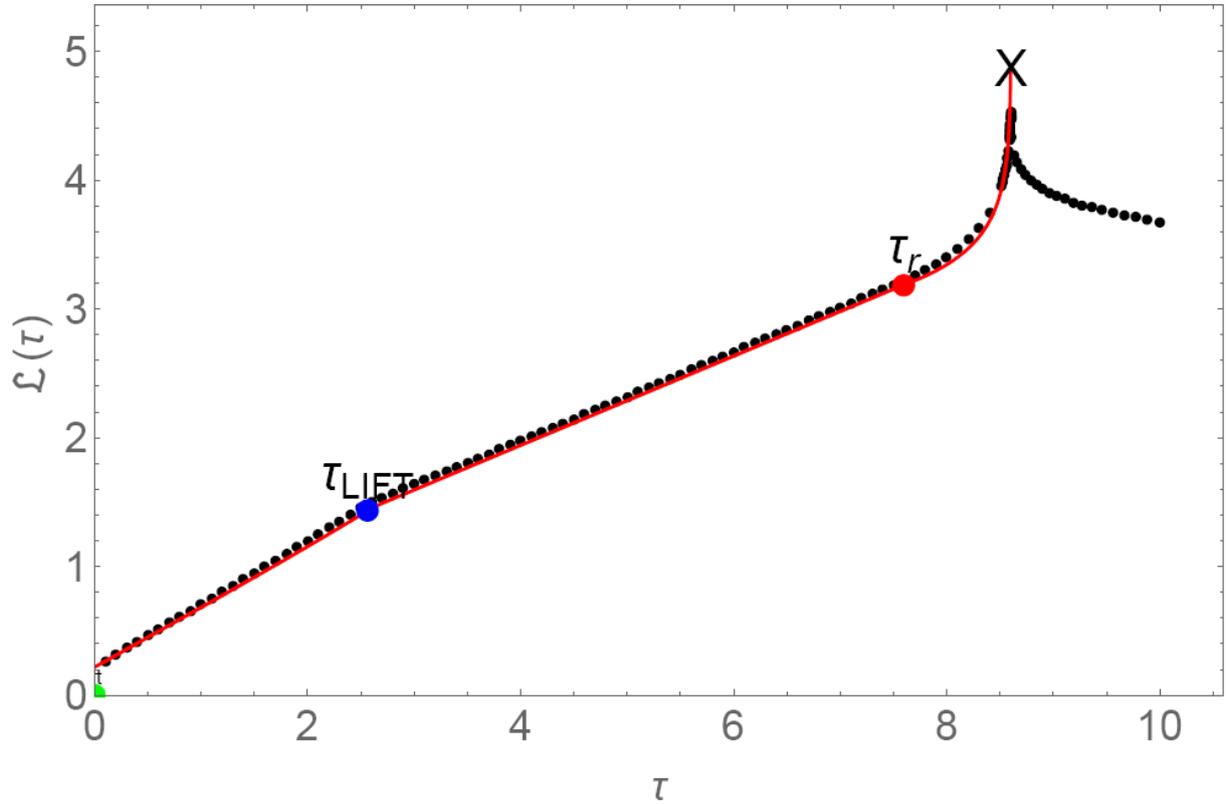

Fig. 6. : Calculated variation of flux integral given by equation 12. The markers $\tau_{LIFT}$ and $\tau_R$ in this and subsequent figures refer to the end of lift-off phase[2] (when the solution parallel to insulator surface reaches the cathode) and the rundown phase (when the solution reaches the end of anode, shown in Fig 1(b)). Points are values of flux integral 12 calculated at discrete values of $\tau$. The solid line (red in online version) is the fit according to equation 13, and the X is the value of $\mathcal{L}(\tau_r+1)$ according to equation 14.

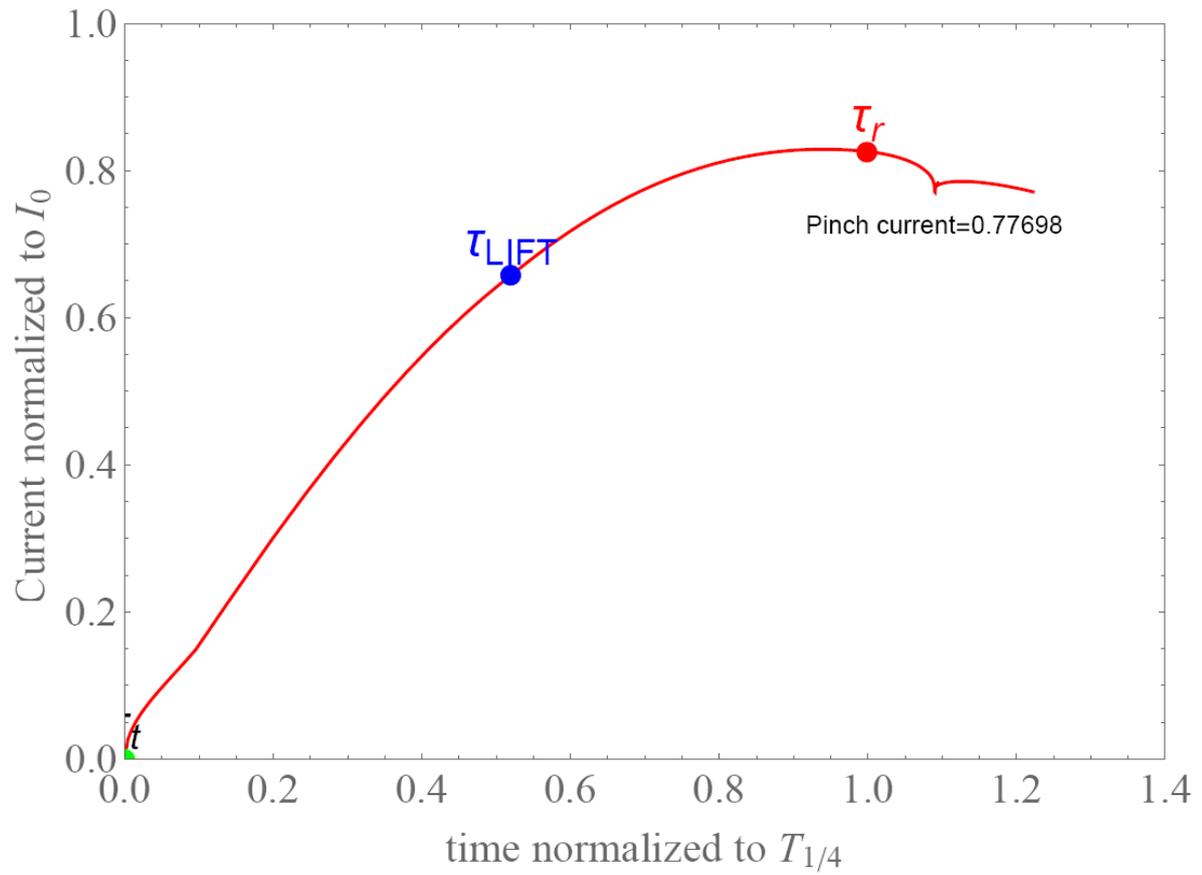

Fig. 7: Current waveform calculated using the method of successive approximation[2] at the third iteration, which exactly coincides the second iteration.

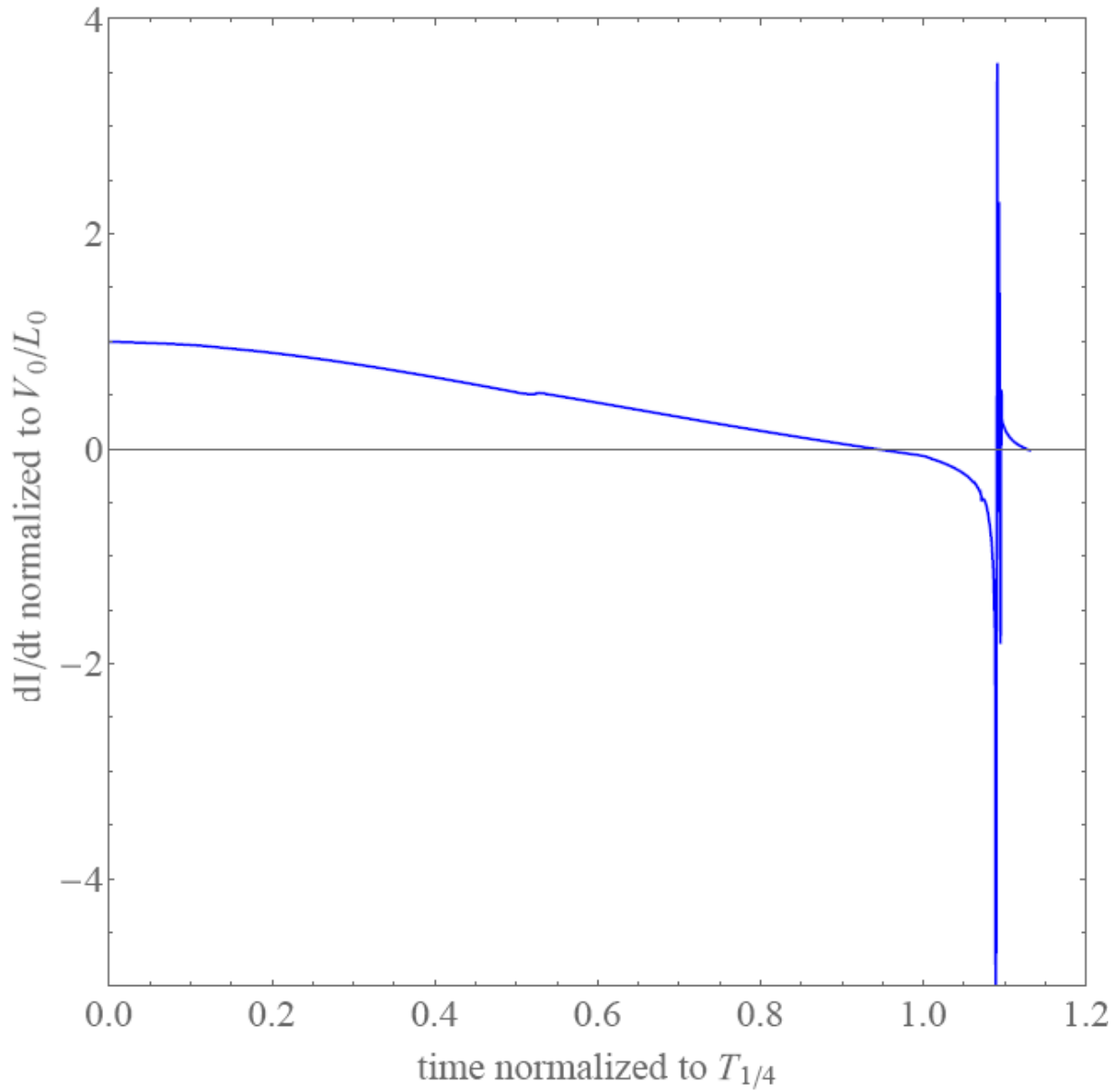

Fig. 8: The normalized current derivative signal.

An important result from this exercise is that the inductance turns out to have a finite value at $\tau_r + 1$. This is not unexpected since the solution of the GV equation for $\tau_r + 1$ has the z-coordinate decreasing to zero just as the radial coordinate goes to zero. The flux depends logarithmically on the radial coordinate but linearly with the axial coordinate; thus the flux integral should reach a finite limit at $\tau_r + 1$. The formula (Equation 39) for inductance given earlier [2] stands revised as given below to specifically include $\tau_r + 1$:

$$\mathcal{L}(\tau) = \tilde{z}_I \text{Log}\left(\sqrt{\tilde{r}_I^2 + \tau}\right) + k_1 \text{Log}(\tilde{r}_C) \tau^{1.5} \quad 0 < \tau \leq \tau_{\text{LIFTOFF}}$$

$$= \mathcal{L}(\tau_{\text{LIFTOFF}}) + \frac{1}{2}(\tau - \tau_{\text{LIFTOFF}}) \text{Log}(\tilde{r}_C) + k_2 \text{Log}(\tilde{r}_C) \quad \tau_{\text{LIFTOFF}} \leq \tau \leq \tau_R \qquad 13$$

$$= \mathcal{L}(\tau_R) - k_3 \text{Log}(\tau_R + 1.00439 - \tau) \quad \tau_R \leq \tau \leq \tau_R + 1$$

$$k_1 = \frac{\lambda_0}{\tilde{r}_c + \lambda_1}; k_2 = \lambda_2 + \lambda_3 \tilde{r}_c + \lambda_4 \tilde{r}_c^2; k_3 = \lambda_5 + \lambda_6 \tilde{r}_c + \lambda_7 \tilde{r}_c^2; \tau_{\text{LIFTOFF}} = \tilde{r}_c^2 - \tilde{r}_I^2; \tau_R = 2(\tilde{z}_A - \tilde{z}_I)$$

$\lambda_0 = 0.276304$; $\lambda_1 = -0.68924$; $\lambda_2 = -0.08367$; $\lambda_3 = 0.105717$; $\lambda_4 = -0.02786$;

$\lambda_5 = -0.05657$; $\lambda_6 = 0.263374$; $\lambda_7 = -0.04005$.

This gives for the inductance at singularity the expression

$$\mathcal{L}(\tau_r + 1) = \tilde{z}_I \text{Log}(\tilde{r}_C) + k_1 \text{Log}(\tilde{r}_C)(\tilde{r}_c^2 - \tilde{r}_I^2)^{1.5}$$
$$+ \frac{1}{2}\left(2(\tilde{z}_A - \tilde{z}_I) - (\tilde{r}_c^2 - \tilde{r}_I^2)\right) \text{Log}(\tilde{r}_C) + k_2 \text{Log}(\tilde{r}_C) + 5.42843 k_3 \qquad 14$$

This is shown in figure 6.

### IV. Summary and conclusions

This paper continues the re-appraisal[2] of the Gratton-Vargas (GV) analytical snowplow model[1] of the plasma focus developed in 1970's to the zone in the vicinity of the DPF axis and to times near the current derivative singularity. This model is a purely kinematic description of what is, in reality, a complicated plasma phenomenon not-quite-well captured even in very sophisticated 3-D Magneto Hydro Dynamic [11], or fully kinetic[12] numerical simulation of plasma processes.

For such an over-simplified model, it is quite effective in fitting observed DPF current waveforms for four large facilities[3], using static inductance, filling pressure and circuit resistance as fitting parameters. Analytical nature of the GV model has been utilized to construct a local curvilinear coordinate system, with unit vectors along the normal to the solution of the GV equation, along the azimuth and along the tangent to the solution, in which conservation laws for mass, momentum and energy can be reduced to effectively-one-dimensional hyperbolic conservation law form[4]. This has been used to demonstrate[4] that axial magnetic field and toroidally streaming fast ions, inferred from experiments over 3 decades of DPF research, are natural consequences of conservation laws in the curved axisymmetric geometry of the DPF. Illustration of its utility as a global parametric optimization tool [5] testifies to its practical importance.

Against this background, this paper attempts a close look at its perceived limitation in the vicinity of the axis at a time close to the current derivative singularity, using prescriptions provided by Gratton and Vargas [1]. In this endeavor, the discussion manages construction of solutions to the GV equation close to the axis for times beyond the singularity. This provides an algorithm for calculation of dynamic inductance of the so called "pinch" phase of DPF. The characteristic signatures of DPF in current and current derivative signals are qualitatively reproduced, *without referring to any plasma phenomena such as instabilities or anomalous resistance or even a hydrodynamic shock wave reflected from the axis*. This is a major new result of this work. It does not imply that these plasma phenomena do not play any role; it merely suggests the possibility that the *occurrence* of current derivative singularity is insensitive to the details of these phenomena; it is quite possible that finer details of the shape of the current derivative contains information specific to some plasma phenomena[6]. Another new result of this exercise is a closed form expression for the dynamic plasma inductance at the moment of singularity. A third new result is the demonstration that the model predicts a power-law dependence of the axial shock-front position on time. Although such power-law dependence is expected from theory of point explosions in ideal gases [13] and has indeed been experimentally observed [14] in DPF, the interesting point about this result is that it comes out of a mathematical model having no connection with hydrodynamics of gases or any other physical phenomena; the only physics contained in the model is the snowplow hypothesis! Also the power-law from the GV model is somewhat different from that observed from the theory, with exponent 0.478 rather than 0.4 from the theory, although this could be related to the fact that what is plotted in Fig. 4 is not real time but the dimensionless time $\tau$. This development follows a very useful suggestion from the Referee, for which the author is extremely grateful.

The scope of this paper is deliberately limited to a discussion of the Gratton-Vargas model; the issue of quantitative comparison of its results with experiments is discussed elsewhere for shortage of space [15]. An important conclusion of such quantitative comparison is that the inductance variation predicted by this paper does not agree with experimental data beyond $\tau = \tau_R + 1$, although the GV model continues to give very good fit to scaled current profiles of 4 large facilities before this instant [15]. This discrepancy is a valuable pointer to understanding DPF phenomenology around the pinch phase in terms of simple models[6].

An unexpected fall-out of this work is the observation of a completely enclosed region, bounded between axially and radially expanding fronts, which naturally occurs in the construction of solution to the GV equation. Its full significance is yet to be understood. It seems quite feasible to construct eigenfunctions of the curl operator (Chandrasekhar-Kendall functions) in a local coordinate system attached to the bound volume, where the theory of Turner Relaxed State[16] can be developed with proper boundary conditions to address recent observations of spontaneous generation of toroidal structures in the neutron mission phase of the PF-1000 DPF experiment [17].